\documentclass[draft]{agujournal2019}
\usepackage{url} 
\usepackage{lineno}
\usepackage[inline]{trackchanges} 
\usepackage{soul}
\usepackage{natbib}
\usepackage{amsmath}
\usepackage{float}

\draftfalse

\journalname{JGR: Space Physics}

\begin{document}
\newcommand{\erf}{\operatorname{erf}}

%
%


\title{Estimation of the electron density from spacecraft potential during high frequency electric field fluctuations}

%
%




\authors{O.W. Roberts \affil{1}, R. Nakamura \affil{1}, K. Torkar\affil{1}, D. B. Graham\affil{2}, D.J. Gershman \affil{3},
J. Holmes \affil{1}, A. Varsani \affil{1},C.P. Escoubet \affil{4}, Z. V\"or\"os \affil{1,5}, S. Wellenzohn \affil{1}, Y. Khotyaintsev \affil{2}, R.E. Ergun\affil{6,7}, B.L. Giles \affil{3}}

\affiliation{1}{Space Research Institute, Austrian Academy of Sciences, Schmiedlstrasse 6, Graz, 8042, Austria}
\affiliation{2}{Swedish Institute of Space Physics, Uppsala, Sweden}

\affiliation{3}{NASA Goddard Space Flight Center, Greenbelt, Maryland 20771, USA}
\affiliation{4}{European Space Research and Technology Centre, Keplerlaan 1, Noordwijk, 2201 AZ, The Netherlands}
\affiliation{5}{Geodetic and Geophysical Institute, Research Centre for Astronomy and Earth Sciences (RCAES), Sopron, Hungary}
\affiliation{6}{Laboratory for Atmospheric and Space Physics, Boulder, CO, USA}
\affiliation{7}{Dept. of Astrophysical and Planetary Sciences, University of Colorado Boulder, Boulder, CO, USA}






\correspondingauthor{O.W. Roberts}{Owen.Roberts@oeaw.ac.at}




\begin{keypoints}
\item Strong AC E-fields can perturb the spacecraft potential by enhancing photoelectron emission making the density estimation unreliable
\item A correction method for the spacecraft potential during large electric fields is presented so that the electron density can be inferred
\item This is tested for three events on the MMS spacecraft showing good agreement with different direct measurements

\end{keypoints}

%
%

%
%


\begin{abstract}
Spacecraft potential has often been used to infer electron density with much higher time resolution than is typically possible with plasma instruments. However, recently two studies by \cite{Torkar2017} and \cite{Graham2018} have shown that external electric fields can also have an effect on the spacecraft potential by enhancing photoelectron escape from the surface. Consequently, should the electron density derived from the spacecraft potential be used during an event with a large electric field, the estimation would be contaminated and the user would see the effects of the electric field rather than density perturbations. The goal of this paper is to propose a method to remove the electric field effects to allow the density derived from spacecraft potential to be used even during large amplitude wave events such as Langmuir waves or upper hybrid waves.  
\end{abstract}

\section*{Plain Language Summary}
Spacecraft in a plasma become charged due to a number of processes. Often the two most important processes in determining the charge are due to the ambient plasma and the photoelectron emission from the surface of a sunlit spacecraft. The potential itself is a function of the electron density, and consequently the potential data can be used to infer the electron density if the photoelectron emission can be modeled. However, in the presence of large electric fields the photoelectron emission can change with the electric field. This means that rather than see fluctuations of density in the spacecraft potential, the effect of the electric field is seen. Here a method is presented to remove the electric field effect on the spacecraft potential such that the density can be estimated even when there are strong electric fields present.

%
%

%


%
%
%
%

\section{Introduction}
Spacecraft embedded in a plasma become charged due to a number of competing processes \citep{Whipple1981}; the spacecraft potential is determined by the balance of several different currents to and from the spacecraft. The size of the currents and their importance for the potential are dependent on several factors; the ambient plasma (ion/electron density and temperature), which gives ion and electron thermal currents denoted $I_{e}$,$I_{i}$. Spacecraft instrumentation (ion/electron emitters) such as electron drift instruments $I_{\text{edi}}$ which emit electrons to measure electric fields \citep{Torbert2016} or active spacecraft potential control (ASPOC)  which regulates the potential by emitting ions $I_{\text{ASPOC}}$ \citep{Torkar2016}. Both of these instruments are present on the Magnetospheric Multiscale Mission spacecraft \citep{Burch2016}  and contribute to the currents when operating.  Ultraviolet radiation which is dominated in near Earth space by Lyman-${\alpha}$ emission from the Sun also causes photoelectron emission $I_{ph}$ from the surfaces of the spacecraft \citep{Kellogg1980,Brace1988}. High energy particles can also cause secondary electron emission $I_{2nd}$ from the spacecraft surface \citep{Pedersen2001,Lai2011}. There is also a small number of backscattered electrons $I_{back}$ \citep{Lai2011}. Finally, there is a small bias current $I_{bias}$ sent from the spacecraft to the electric field probes with the purpose of keeping the probe potentials close to zero potential with respect to the plasma.

Under certain conditions the contributions of many of these currents can be neglected. In a tenuous plasma the contribution of the ion thermal current can be neglected \citep{Lybekk2012}. The secondary electron yield from a spacecraft depends on the material of the spacecraft and the energy of the primary electrons. The peak of the yield is generally near 300-800eV \cite[e.g.][]{Lai2011,Balcon2012}, however should the electron temperature be outside the peak yield the effect is mitigated. Therefore should the electron temperatures be favourable the contribution of secondary emission can also be neglected. Backscattered electrons can also contribute to the currents and have similar energies to the incident electrons \cite[e.g.][]{Darlington1972,Sternglass1954}. However the probability of a backscattered electron occurring is much lower than for a secondary electron. If secondary electrons can be neglected then so can backscattered electrons. The bias current is selected to be of the order of $\sim-100$nA for each of the six electric field probes on MMS. This gives a combined bias current of $\sim-600$nA on one of the spacecraft which is much smaller than the other current sources and can be neglected. Therefore if the plasma is sparse with low temperatures and both ASPOC and EDI are not operating the photoelectron current from the spacecraft and the electron thermal current to the spacecraft dominate and are approximately equal in magnitude $I_{e}+I_{\text{phot}}\sim0$.

The electron thermal current can be approximated by Eq. \ref{Eq2} by assuming that the particles have a Maxwellian velocity distribution function \citep{Mott-Smith1926,Pedersen1995}:

\begin{equation}
    I_{e}=-A_{\text{spac}}qn_{e}\sqrt{\frac{k_{B} T_{e}}{2 m_{e} \pi}} \left(1+\frac{q  V_{sc}}{k_{B}T_{e}}\right)
    \label{Eq2}
\end{equation}

Where $A_{\text{spac}}$ is the spacecraft's surface area. The MMS spacecraft have an approximate area of 34m$^{2}$. This value is given in the MMS technical report \citep{Davis2009} and is the value typically used in the literature \cite[e.g.][]{Torkar2016,Andriopoulou2016,Torkar2017,Nakamura2017,Andriopoulou2018,Graham2018,Torkar2019}. There may be some small difference between the value of the technical report and the precise value which may give a small systematic error in the determination of the photoelectron parameters.
The photoelectron current can also be approximated by assuming several (usually two) populations of photoelectrons and performing a fit of the electron thermal current (Eq. \ref{Eq2}) to the spacecraft potential.  

\begin{equation}
    I_{\text{phot}}=I_{\text{ph}0}\exp{\left(-\frac{qV_{sc}}{k_{B}T_{ph0}}\right)}+I_{\text{ph}1}\exp{\left(-\frac{qV_{sc}}{k_{B}T_{ph1}}\right)}
    \label{Eq3}
\end{equation}

Here, $I_{\text{phot}0,1}$ and $T_{ph0,1}$ are fitted parameters determined later in Fig \ref{fig:2}.  Under the conditions when all other sources of current are small and an assumption (or a direct measurement if possible) is made about the ambient electron temperature, the spacecraft potential can be calibrated to give a measurement of the electron density (\citealt{Pedersen1995},\citealt{Escoubet1997},\citealt{Nakagawa2000},\citealt{Pedersen2001},\citealt{Pedersen2008}) given in Eq \ref{neest}.

\begin{equation}
    n_{e,SC}=\frac{1}{qA_{\text{spac}}} \sqrt{\left(\frac{2\pi m_{e}}{k_{B}T_{e}}\right)}\left(1+\frac{qV_{sc}}{k_{B}T_{e}}\right)^{-1}\left(I_{ph0}\exp\left({\frac{-qV_{sc}}{k_{B}T_{ph0}}}\right)+I_{ph1}\exp\left({\frac{-qV_{sc}}{k_{B}T_{ph1}}}\right)\right)
    \label{neest}
\end{equation}

It is often the case that plasma instruments mounted on a spinning spacecraft use the spin to sample different directions thus limiting the sampling rate to the spacecraft spin frequency. Using the spacecraft potential overcomes this limitation and allows for a much higher time resolution than is typically possible for plasma instruments. This makes the electron density deduced from the spacecraft potential useful for investigating kinetic scale density fluctuations (\citealt{Kellogg2005}, \citealt{Yao2011},\citealt{Chen2012b}, \citealt{Roberts2017}, \citealt{Roberts2018}, \citealt{Roberts2018b}), or for use when direct measurements are not available \citep{Haaland2009}.

However there are limitations for the use of the spacecraft potential as a measurement for the density. The first limitation is that the illuminated area of the spacecraft changes as it is rotating, resulting in the potential being a function of the spacecraft phase angle. This causes several peaks to appear in the Fourier spectrum of the spacecraft potential (or the density fluctuations derived from it) \citep[e.g.][]{Kellogg2005}, which need to be removed by bandpass filtering \citep{Yao2011}, subtracting harmonics \citep{Torkar2017} or developing an empirical model of the spacecraft charging \citep{Roberts2017}. Another limitation is that the spacecraft have a finite charging time \cite[e.g.][]{Chen2012b}, therefore it will take some time for the spacecraft to respond to an external driver and reach a new equilibrium potential. This timescale is typically very small (as will be discussed later) however it can affect the potential if an abrupt change occurs such as a high frequency wave.

The other limitations occur for two reasons; either other currents are present such that the relation $I_{e}+I_{\text{phot}}\sim0$ is no longer valid, or that the assumptions of Eqs. \ref{Eq2} or \ref{Eq3} are no longer valid. Densities can be inferred when the ASPOC instrument is on, as the current itself is known. However it is typically more difficult to infer the density as the variations in spacecraft potential due to the density are smaller as the potential is regulated (\citealt{Torkar2015}, \citealt{Andriopoulou2015}, \citealt{Andriopoulou2016}, \citealt{Nakamura2017}, \citealt{Andriopoulou2018}). Meanwhile, changes due to the electric field as described in the following paragraph may be comparable to when ASPOC is off. Additionally, regions such as the magnetosheath contain particles which contribute to a secondary emission current \citep{Pedersen2001}.   

More recently another limitation has been discovered, where the electric field can have an effect on the spacecraft potential in modifying the photoelectron escape from the spacecraft. It was shown that in strong DC and AC electric fields can have an influence on the spacecraft potential \citep{Mcfadden2003,Malaspina2014,Torkar2017,Graham2018}. \cite{Graham2018} used data from MMS to show that when high frequency upper hybrid, or Langmuir waves are present (such as are common near the Earth's magnetopause \citep{Graham2018a}) they enhance photoelectron escape from the spacecraft leading to an increase in the spacecraft potential. This conclusion was supported by numerical simulations that showed that the presence of an electric field causes enhanced photoelectron escape. Physically photoelectrons can have a variety of energies and not all photoelectrons escape the potential well of the spacecraft. Some photoelectrons are not sufficiently energetic to escape and return to the spacecraft. The presence of a large electric field would cause some photoelectrons with lower energies, which ordinarily would have returned to the spacecraft to escape. This results in a measurable change in the spacecraft potential.
By using the high time resolution electric fields, spacecraft potential and the direct electron measurement it was concluded that the observed increase in spacecraft potential was not associated with density perturbations. 

There are several physical applications where high time resolution density measurements are desirable during strong electric fields. The purpose of this paper to determine the effect that the electric field has and to remove it leaving a spacecraft potential value, which can be used to derive the density or density fluctuation in such an event. Some examples where this technique might be useful when investigating increments of the density, or when evaluating the Pondermotive force which requires a measurement of both density and electric field \citep{Henri2011}. Electrostatic signatures of PSBL mixing such as phase-space holes and double layers have significant implications for energy transport in the magnetosphere. Our best measurements of 3D electron holes on MMS are in the PSBL/Lobe boundary \cite[e.g.][]{Tong2018,Holmes2018,Mozer2018} on millisecond timescales with densities of $\sim 0.1$ cm$^{-3}$. Waves are common in this region, and having increased accuracy in electron density measurements would be highly beneficial in interpreting these structures. Other applications may include investigating the effects of density perturbations on wave propagation \citep[e.g.][]{Moullard2002,Yearby2011}. Although to test and present the method here we investigate simpler cases to demonstrate the concept of the correction method. 

This work will present an empirical method to remove the perturbation in the spacecraft potential due to high frequency waves. In the cases which will be presented the density doesn't vary strongly which is desirable for testing the method. While the correction is not specifically important for the cases studies here, testing on quiet intervals is a necessary first step before the correction is applied for more complex intervals. In the following section we will present the data used from the MMS spacecraft, which were initially characterized by \citealt{Graham2018},\citealt{Graham2018a}. In the third section we will present a method to remove the electric field contribution from the potential such that we are left only with the part which is due predominantly to electron density perturbations.

\section{Overview of the Events}
In this section an overview of three events will be presented. The three events consist of a variety of different electron densities and spacecraft potentials. For the first event we will present the technique step by step and for the other two intervals only the results will be presented. In all of the intervals burst mode data were available, and AC coupled electric field measurements sampled at 65536 Hz (termed HMFE) are also available. In Fig. 1 several measured parameters for the first interval from the MMS2 spacecraft \citep{Burch2016} are presented.  This is a one second interval starting at 2016-10-16 20:33:40.500 UT. Figure 1(a) shows the high frequency AC coupled electric field measurements from the Spin Plane Double Probes instrument \citep{Lindqvist2016} and the Axial Double Probes \citep{Ergun2016}. These are presented in a co-ordinate system which is aligned with the mean magnetic field direction defined as the average measured from the fluxgate magnetometer \citep{Russell2016} during the interval. The direct ion and electron measurements are obtained from the Fast Plasma Investigation's (FPI) Dual Ion Spectrometer (DIS) and Dual Electron Spectrometer (DES) \citep{Pollock2016} which have sampling rates of 6.6 Hz and 33.3Hz respectively.

This interval was investigated by \cite{Graham2018},\cite{Graham2018a} and identified as an example of a Langmuir wave  with $E_{\parallel}\gg E_{\perp}$ which causes the electron density estimation from spacecraft potential to become inaccurate. Figure 1(b) shows the envelope of the total fluctuations. For a signal $S$ this is obtained by finding the analytic signal from the Hilbert transform $H(S)$ and summing up their magnitudes $A_{i}=\sqrt{S_{i}^2+H(S_{i})^2}$. This is done for each component and then gives the magnitude of the envelope $E_{\text{env}}=\sqrt{A_{x}^2+A_{y}^2+A_{z}^2}$. Figure 1(c) shows the spacecraft potential measured during the event with a sampling frequency of 8192Hz, which has a striking similarity to the electric field envelope fluctuations. The spacecraft potential is calculated using the mean of the four biased spin plane probes $V_{j=1,2,3,4}$ to calculate the probe to spacecraft potential $V_{psp}$. Each probe is given a bias current such that is selected to approximately balance the electron thermal and photoelectron currents from the probe. This ensures that the the probe has a positive value close to zero for the potential with respect to the plasma.  If one probe is disabled such as is the case with MMS4 when a bias current failure occurred on 2016-06-12 05:28:48 then the mean of two opposing probes is used. To obtain the spacecraft potential the probe to spacecraft potential is used converted to a spacecraft potential using;

\begin{equation}
    V_{sc,i}=1.2(-V_{psp})+c_{i}
    \label{potentialDef}
\end{equation}

where $V_{psp}$ is:

\begin{equation}
    V_{psp}=\frac{1}{N}\sum_{j=1}^{j=N} V_{\mathrm{probe},j}
\end{equation}

where the coefficient of 1.2 in Eq \ref{potentialDef} is a shortening factor to account for the fact that the probes are not infinitely away from the spacecraft and $[c1,c2,c3,c4]=[1.3, 1.5, 1.2, 0.0]$V are constants (i.e. the nominal probe to plasma potential) which are determined from the photoelectron energies seen in FPI. It is important to note that the probes are themselves not strictly at zero potential with respect to the local plasma, and there can be some error in the derivation of the nominal probe to plasma potential that affects the determination of the photocurve. However, this error is likely to be small and would also only cause a systematic error \citep{Torkar2015,Andriopoulou2015}. 

One final point to note is that MMS is a spinning spacecraft and the potential and electric field which are both required for this analysis is calculated from measurements of the booms in the spin plane (with a third component of the electric field coming from the axial double probe). Perfect electric field measurements by probes rely on identical probe-to-plasma potentials on all probes, which requires identical illumination and plasma conditions. Therefore, when an electric field probe is shadowed and in the wake behind the spacecraft, there will be a spacecraft generated electric field which scales with the potential. The SDP probes are mounted on the spacecraft at the spin phase angles of $[30,120,210,300]^{\circ}$ \citep{Lindqvist2016}. Where $0^{\circ}$ is defined as having the $x$ component of the spacecraft Body coordinate system pointing sunward. Shadowing of a probe occurs at these phase angles where a boom is sun-aligned. These brief events occur at every quarter of a spacecraft spin period of 20 s and lie outside the analyzed intervals as is detailed in Fig \ref{fig:spineffect} later.

A cross correlation of the electric field envelope (resampled to the spacecraft potential time tags) and the spacecraft potential shows a maximum correlation of 0.96 and a corresponding time lag of 0.0024s with the electric field leading the spacecraft potential. In all intervals ASPOC is off, and EDI is in passive mode (i.e. the electron gun is off) allowing us to use equations \ref{Eq2}-\ref{Eq3} to be used to derive the electron density.


\begin{figure}[H]
\includegraphics[width=0.5\textwidth]{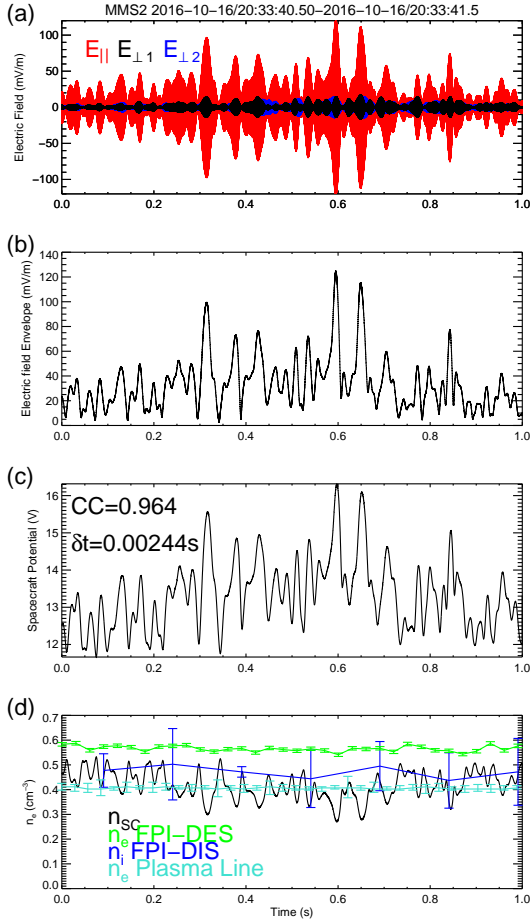}
\caption{Overview of the first event 2016-10-16 20:33:40.50-20:33:41.5 from MMS2 (a) Shows the three components of the electric field fluctuations in field aligned coordinates. (b) The electric field envelope obtained from the Hilbert transform. (c) The measured spacecraft potential, (d) The estimation of the density from the spacecraft potential in (d) is shown in black and the direct measurement from FPI-DES along with the error bars. The FPI error bars denote the statistical uncertainty in the FPI measurement while the error bars on the plasma line denote 1/100 th of the peak power.}
\label{Lab:Fig1}
\end{figure}

\begin{figure}
    
    \includegraphics[angle=-90,width=0.5\textwidth]{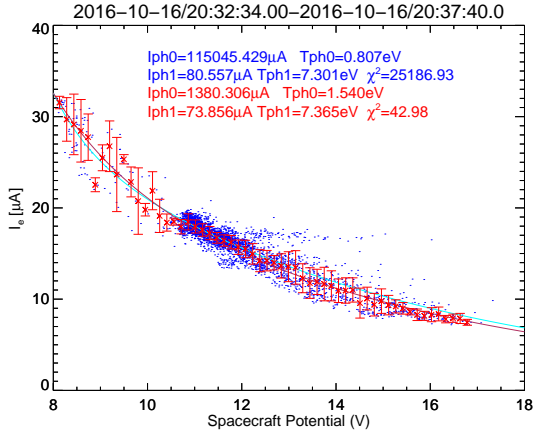}\\
  
    \caption{The photocurve used to determine the photoelectron parameters. The measured points are given in blue and the binned data are shown in red with the corresponding error bars in red. Fits are performed to the unbinned data in blue and to the binned data in red.}
    \label{fig:2}
\end{figure}

The values for the photoelectron parameters are obtained by fitting two exponential curves (Eq. \ref{Eq3}) to the data in Fig \ref{fig:2}. To ensure a large enough range of potentials the curve needs to be determined over a longer time period than the one second interval of interest presented in Fig \ref{Lab:Fig1}, where two burst mode intervals are used between 20:32:40-20:33:44 and 20:35:24-20:37:04. This choice is motivated by the need to have a large enough interval so that a range of spacecraft potentials are sampled but also small enough so that there are no changes in the photoelectron parameters e.g. a change in electron temperature/ Solar UV flux. The electron thermal current is determined from Eq. \ref{Eq2} from the FPI-DES electron measurements and the spacecraft potential measurements. As there is a large spread in the data (shown in blue); they are binned to limit the effects of outliers (and regions with large electric fields), and the median $p_{i}$ in each bin and the error $\sigma_{i}$ given by Eqs \ref{median}-\ref{error} are shown in red; 

\begin{equation}
    p_{i}=\text{median}(I_{e,i})
    \label{median}
\end{equation}

\begin{equation}
    \sigma_{i}= \frac{1}{\sqrt{2}\erf^{-1}(1/2)}\text{median}(\lvert I_{e,i}-p_{i}\rvert)
    \label{error}
\end{equation}

where the error is assumed to be Gaussian. The fitting is performed with both the binned (red) and the unbinned points (blue) and although the $\chi^2$ value is large for the unbinned values, both the binned and unbinned data yield similar results, we will use the values based on the binned data. The fitted parameters are also indicated on the Fig \ref{fig:2}. Two populations of photoelectrons are required as the photocurve is often characterised by several different exponential laws at different values of the potential. Typically exponential laws are fitted between 0-4V, 4-9V and $>$9V \cite[e.g.][]{Pedersen1995, Andriopoulou2015,Andriopoulou2018}. The ranges of the potentials here span two of these regions necessitating a dual fit. However as there is only a small region below 9V the parameters with subscript 0 are poorly estimated. To estimate the total photoelectron current the curve would need to be extrapolated to 0V which is impossible to do meaningfully as there are no data points below 8V. A reasonable value for the total photoelectron current density of all populations is $70\mu$ A m$^{-2}$ \cite{Davis2012}. The highest energy population of photoelectrons here gives a current density of $12 \mu$ A m$^{-2}$ (assuming a projected sunlit area of 5.9m$^{2}$) which is slightly larger than what is measured for this potential range typically for MMS  $3.4 \mu$ A m$^{-2}$ \citep{Andriopoulou2018}. The energy of this population is also smaller 7.4V versus 12.1V in \cite{Andriopoulou2018}. In \cite{Andriopoulou2018} the photoelectron curve was calculated only when electric fields were less than 10mV/m. This reflects that the interval has somewhat higher photoelectron emission with respect to times when the electric field is low.   

The estimation of the electron density from the uncorrected spacecraft potential using these parameters is shown in Fig. 1(d) and the corresponding direct measurement of ion and electron density from the FPI. This procedure is performed for all the intervals, and a separate photocurve is derived for each event. A discrepancy between the electron density measurements and the estimation from the spacecraft potential has been observed when large electric field fluctuations are present as opposed to regions where the amplitude of the electric field is smaller \citep{Graham2018}.

For an additional estimate of the electron density a short time Fourier transform is applied on the $E_{\parallel}$ component of the electric field with a window length of 2048 points ($0.0625$ times the Nyquist frequency) and zero overlap is used. The Langmuir wave has the dispersion relation given in Eq. \ref{LangmuirDispersion}. The peak frequency corresponds to an electron density of $n_{e}=\frac{m_{e}\epsilon_{0} 4\pi^{2} f_{pe}^2}{q^2}$ where the electron pressure (rightmost term of Eq \ref{LangmuirDispersion}) is ignored. This peak is extremely sharp and well defined and the error bars indicate the width of the peak at one hundredth of the maximum power. The estimates from the plasma line and the FPI-DES in Fig \ref{Lab:Fig1} significantly differ. For validation of the spacecraft potential correction and the final electron density estimation, comparison with a direct measurement is required. This is more difficult when the two other measurements disagree by about 0.2cm$^{-3}$. Note that this is much larger than the statistical uncertainty in the FPI electron density measurement. The discrepancy between the two different direct methods of estimating the density are therefore significant and are discussed in the next subsection.

\begin{equation}
    \omega^2=\frac{n_{e}e^2}{m_{e}\epsilon_{0}}+3k^2\left(\frac{k_{B}T_{e}}{m_{e}}\right)
    \label{LangmuirDispersion}
\end{equation}

\begin{figure}[H]
\includegraphics[width=0.5\textwidth]{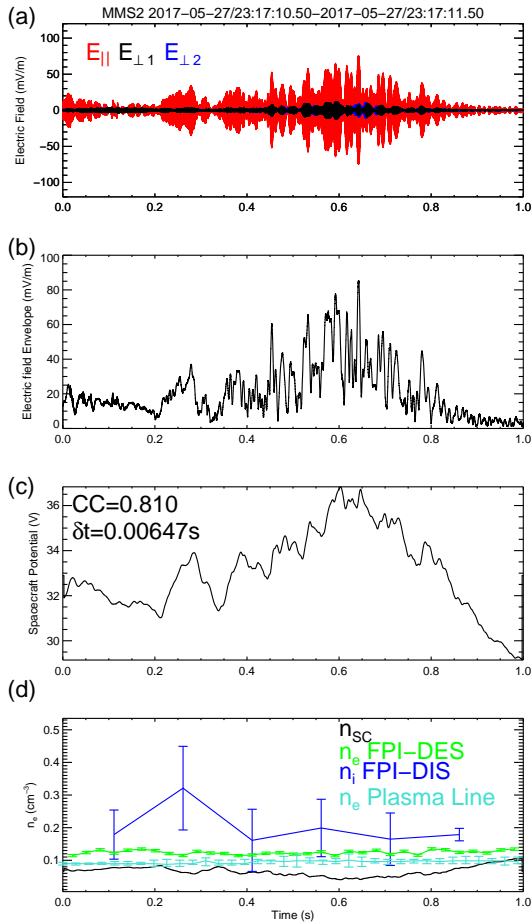}
\caption{Overview of the second event 2017-05-27 23:17:10.50-23:17:11.50 from MMS2. The panels (a)-(d) denote the same as Fig. \ref{Lab:Fig1}}
\label{Over2}
\end{figure}

\begin{figure}[H]
\includegraphics[width=0.5\textwidth]{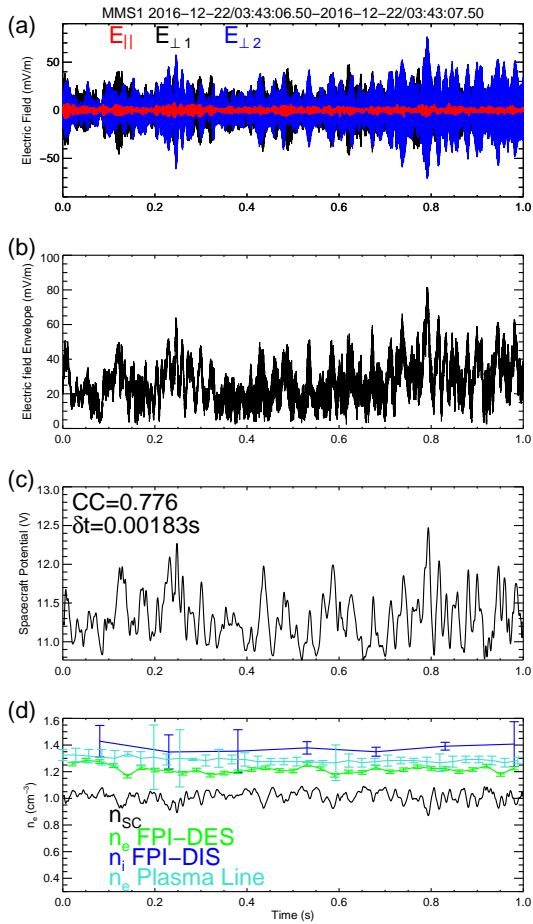}
\caption{Overview of the third event 2016-12-22 03:43:06.50-03:43:07.50 from MMS1. The panels (a)-(d) denote the same as Fig. \ref{Lab:Fig1}}
\label{Over3}
\end{figure}

Two other intervals are presented in Figures \ref{Over2} and \ref{Over3}, which have the same format as Fig. \ref{Lab:Fig1}. The second interval occurs on MMS2 between 2017-05-27 23:17:10.50-23:17:11.50. This interval is of a much sparser plasma $(n_{e}<0.2$cm$^{-3}$) compared to the first interval where the spacecraft are in the plasma sheet boundary layer (PSBL) where the spacecraft potential is very large $(V_{sc}>28V)$. Similarly to the previous event there is a Langmuir wave present. The correlation coefficient between the spacecraft potential and the electric field is also large (CC=0.81) and the lag is longer due to the longer time constant for discharging in thin plasma, and the larger potential in this environment leads to longer charging times. The final event presented has a much higher density and a lower spacecraft potential. In contrast, this event is dominated by an upper hybrid wave which is characterized by the dominance of the perpendicular components of electric field ($E_{\perp} \gg E_{\parallel}$). Moreover, several electron Bernstein waves above and below the upper hybrid frequency are present. This results in a more complicated electric field envelope with rapid fluctuations at several frequencies. The power of the Upper hybrid wave dominates the Fourier spectra and the density from the plasma line is estimated from the dispersion relation for the upper hybrid wave in Eq \ref{UHwave}, where the correction due to the electron thermal speed is ignored, and magnetic field is obtained from the fluxgate magnetometer.

\begin{equation}
   \omega^2=\frac{n_{e}e^2}{m_{e}\epsilon_{0}}+\left(\frac{qB}{m_{e}}\right)^2+3k^2\left(\frac{k_{B}T_{e}}{m_{e}}\right)
   \label{UHwave}
\end{equation}

\subsection{Estimation of the electron density}

In this section the two different measurements of the electron density that will be used for validation are discussed.  Firstly the direct measurement from FPI-DES will be discussed. The limitation of this measurement is that the instrument also measures photoelectrons which originate from the surface of the spacecraft (external), or within the instrument itself (internal) \cite[e.g.][]{Gershman2017}. This can result in a higher measured density than the true value of the ambient plasma especially in a sparse plasma where the counts are much lower and the photoelectrons may contribute a larger fraction to the final measurement. The FPI moments calculations remove any electrons channels below $q V_{sc}$ excluding any external photoelectrons. However, internal photoelectrons can also be present in the data and can have energies exceeding $q V_{sc}$ \citep{Gershman2017}. To correct for this, in the moments data on the MMS archive, a model is used \citep{Gershman2017} but some differences may be present in each individual case. 

\begin{figure}[H]
\includegraphics[width=0.85\textwidth]{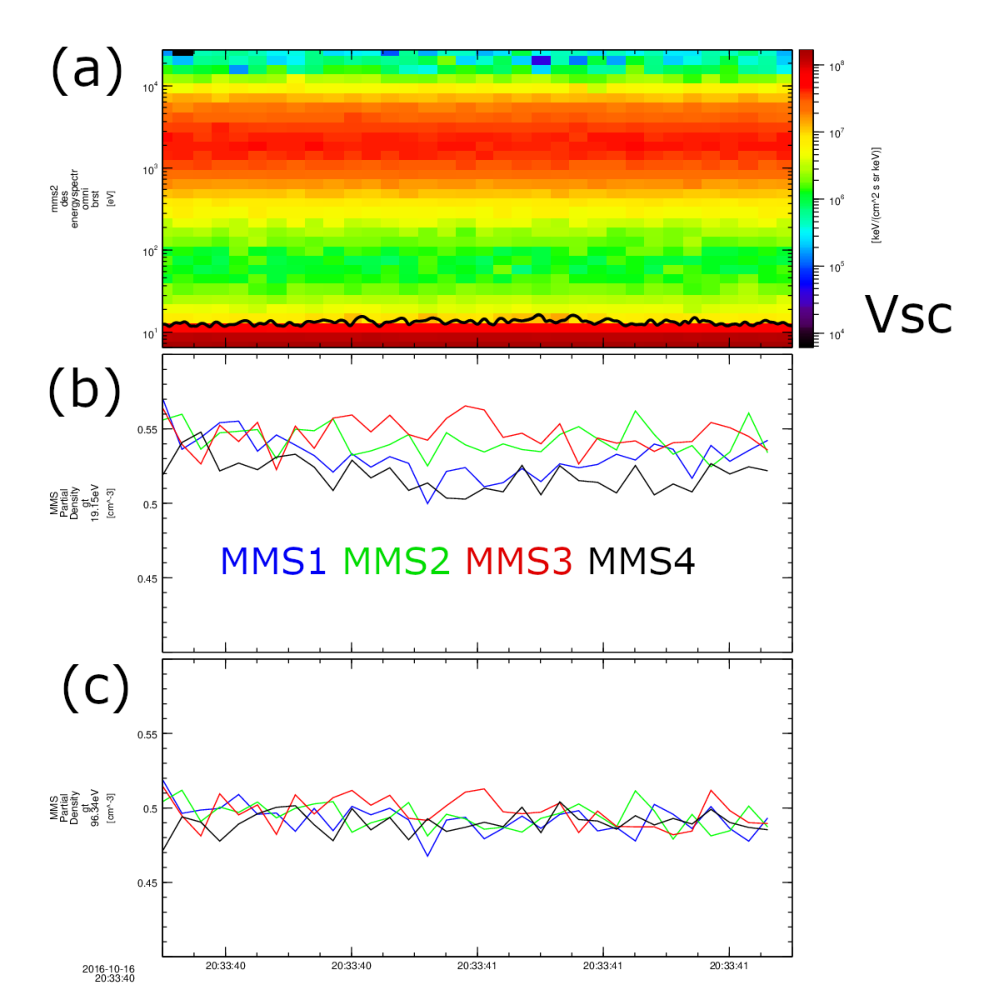}
\caption{Overview of the electron and spacecraft potential measurements on interval 1 (a) Omnidirectional electron spectra with the spacecraft potential overplotted, (b) partial density moments with starting integration energy of 19.6eV (c) and with starting integration energy of 96eV.}
\label{fig:photoelectrons}
\end{figure}

To investigate further we use the partial moments data set of the FPI-DES data, where the starting energy of the integration is at two different energies. Figure \ref{fig:photoelectrons}a shows the omnidirectional electron energy spectra. The black curve in Fig \ref{fig:photoelectrons}a denotes the spacecraft potential. In the region where the energies of the particles are smaller than $q V_{sc}$, higher counts can be seen in the spectrogram. These are external photoelectrons. However some photoelectrons are also seen at higher energies exceeding $q V_{sc}$ although it is less intense. These are likely internal photoelectrons which can have higher energies than the spacecraft photoelectrons \citep{Gershman2017}. 

   Figure  \ref{fig:photoelectrons} (b) has a starting integration energy of 19.6eV, in principle this should not contain any spacecraft photoelectrons. The ion inertial length in this case is $\sim 300km$ while the electron inertial length is of the order of $\sim <7$km. When compared to the inter-spacecraft distances which are also $\sim 7$km we would expect similar densities to be observed on each spacecraft. However there seems to be a systematic difference between the spacecraft when the moments are summed for energies greater than 19.6eV. In figure \ref{fig:photoelectrons}(c) the starting integration is set to 96eV. In this figure the densities between different spacecraft agree much better with one another. This suggests that the differences between the spacecraft in Fig \ref{fig:photoelectrons}(b) are due to instrument photoelectrons. Furthermore the mean value is slightly reduced by roughly 0.1cm$^{-3}$ giving better agreement with the plasma line. 

The other measurement of the electron density comes from the plasma line. While this is useful for comparison it is limited to times when the plasma wave is sharp and well defined. The dispersion relation of the Langmuir wave was given by Eq \ref{LangmuirDispersion}. In deriving the density from the plasma line effects due to electron thermal speed (second term on the right hand side of Eq. \ref{LangmuirDispersion}) have been neglected. This could be a potential source of error on the plasma line density estimation which shall be investigated now.  


 To test whether this is the case the second term is calculated using the measured value of the electron temperature from FPI for a Langmuir wave with $k=\frac{0.1}{\lambda_{D}}$. The electron temperature in this case is $T_{e} =1.04\pm 0.02$keV, and the Debye length  $\lambda_{D}=319\pm3$m. The value of the correction term is two orders of magnitude less than the first term in Eq. \ref{LangmuirDispersion}. Another potential source of error comes from the Doppler shift of the wave into the plasma frame $\omega_{pla}=\omega_{sc}-\mathbf{k}\cdot\mathbf{v}$. While this is an important correction for low frequency waves such as those often observed in the solar wind \cite[e.g.][]{Roberts2013} it is less important for high frequency waves. Here the $\mathbf{k}\cdot\mathbf{v}$ term is several orders of magnitude less than the wave frequency when using either ion or electron bulk speeds. This suggests that the most accurate measurement of the electron density for this case comes from the plasma line. The plasma line has been used for cross calibration on the Cluster mission  as an active sounding instrument is present \cite[e.g.][]{Fazakerley2010,Trotignon2010}. However, no such instrument is present on MMS and ideal conditions for estimating the density from the plasma line may not always be present which motivates the development of using the spacecraft potential as an additional measurement of density.

For validation of the electron density estimation from spacecraft potential both the plasma line and the partial moments with integration starting at 96eV for intervals 1 and 2 will be used. As the density is much higher and photoelectrons relatively less important 14eV will be used as the starting energy for integration in interval 3.

\section{Method}

\begin{figure}
    \centering
    \includegraphics[width=0.5\textwidth]{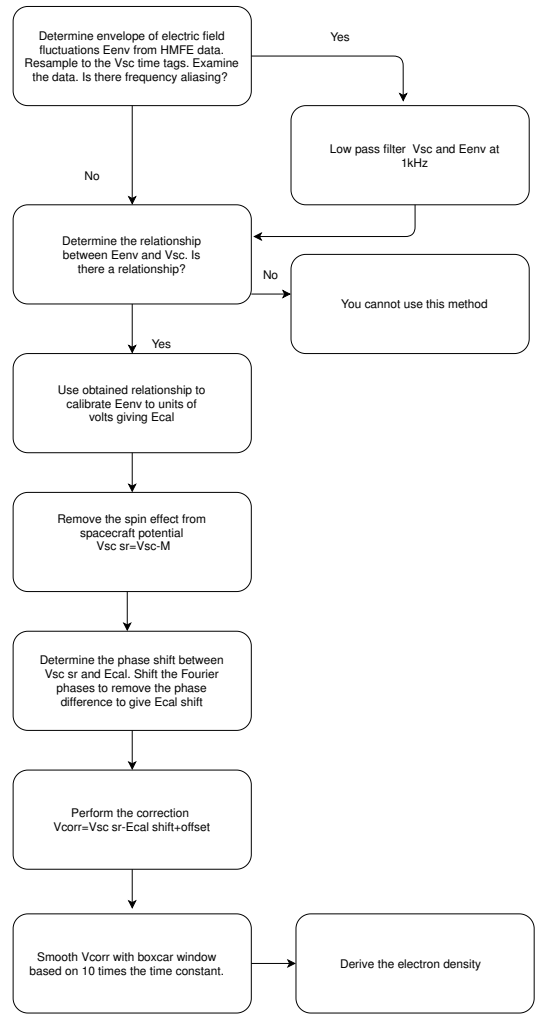}
    \caption{Flowchart describing the method}
    \label{fig:flowchart}
\end{figure}

The principle of this method is to use the envelope of measured electric field fluctuations $E_{\text{env}}$ obtained from the SDP and ADP instruments and calibrate the electric field in units of mV/m to a corresponding spacecraft potential change in units of V. The contribution of the electric field to the spacecraft potential can then be removed, in principle leaving only the perturbations due to electron density and temperature. This method allows the spacecraft potential density estimate to be used under certain conditions when there are large electric fields present. A figure summarizing the method is given in Fig \ref{fig:flowchart}. Firstly the data should be inspected to determine whether it is suitable. To be able to apply this method there should be a clear relationship between $E_{\text{env}}$ and $V_{\text{sc}}$. Should that not be the case then the contribution of the electric field envelope to the change in potential cannot be accurately determined and the method cannot be used.

\begin{figure}
    \centering
    \includegraphics[angle=-90,width=0.5\textwidth]{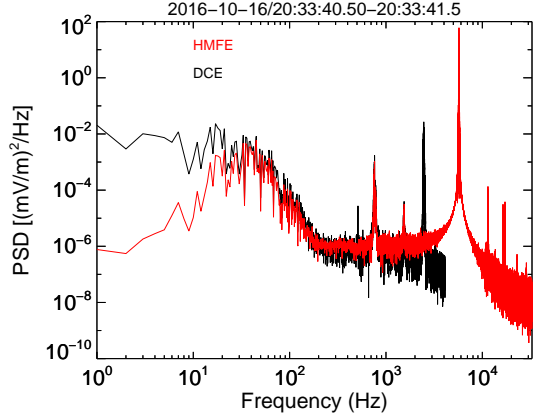}
    \caption{Fourier power spectra of the trace DC electric field fluctuations (black) and the trace HMFE electric field fluctuations (red)}
    \label{HMFE}
\end{figure}

The spacecraft potential time series is sampled at a lower rate compared with the HMFE data therefore there may be effects of frequency aliasing which may need to be removed by low pass filtering. This low pass filtering will need to be done on both the spacecraft potential and the re-sampled electric field envelope data. This is dependant on the frequency of the plasma waves and will vary from case to case. In the first case some aliasing can be seen in the high frequency coherence (Figure \ref{fig:Fig7}, and in the electric field data in Fig\ref{HMFE}), however in the first interval this does not need to be performed as the power of these fluctuations in the spacecraft potential is low. In the second and third cases low pass filtering does need to be performed. 

In figure \ref{HMFE} we compare the Fourier power spectra of both the HMFE data and the DCE. The two different measurements differ by their sampling rates, furthermore the HMFE data are high pass filtered at 100Hz. The effect of aliasing can also be seen here in the DCE data where there is an aliased peak near 2000Hz. It is important to note that the photoelectron emission due to electric field waves is a function of both the wave frequency and amplitude \citep{Graham2018}. Lower frequency waves cause more photoemission as do higher amplitude waves. In this work we consider the removal of only high frequency waves and make an assumption that the low frequency waves do not affect the photoemission significantly. For the intervals concerned the high frequency waves have much larger amplitudes than lower frequency Fourier modes. This is shown in fig \ref{HMFE} where the Fourier modes associated with the Langmuir wave have power which is approximately four orders of magnitude larger than any other Fourier mode. Therefore we assume that the enhanced photoemission associated with high frequency wave components are dominant over any low frequency components. 

The next step is to calibrate the electric field $E_{\text{env}}$ to units of volts to give $E_{\text{cal}}$, figure \ref{fig:fig6} shows the calibration curve for the first event. We use the same procedure to bin the data as was done for the photocurve in Fig \ref{fig:2}, and fit both a linear and a second order polynomial to the data. Both curves fit well but the second order polynomial fits better at both low and high spacecraft potentials and has a smaller $\chi^{2}$ value, and give a different value of the background potential which differs by $\sim 0.5V$. This may lead to difficulties in determining the value of the 'background' spacecraft potential meaning there might be an offset in the electron density estimation. For all cases the second order polynomial fit has a smaller $\chi^{2}$ value and will be used calibrate the electric field. This curve agrees well with the simulations shown in Fig 8a of \cite{Graham2018}, the change in potential in the simulation was found to be consistent with enhanced emission of photoelectrons. This is caused by photoelectrons which would have returned to the spacecraft being accelerated away from the spacecraft by the electric field. An alternative hypothesis is that high energy particles trapped in the wave could cause increased secondary electron emission. This effect would likely be small as the increased electron temperature would cause the electron thermal current to increase. Any increase in currents due to secondary electron emission would likely be small compared to the electron thermal current. During the three intervals studied the electron temperatures are 1046eV, 185eV and 127eV respectively which are all outside of the peak in secondary emission yield. Furthermore, these temperatures are similar in the 1 second interval before and after the intervals studied here suggesting that trapped high energy electrons are not the cause of the potential changes here.

In addition to the effect of the electric field there is an effect of the spin from the spacecraft, and the amount of illuminated surface changing throughout a spin. This is due to different areas being exposed to sunlight, because of the geometry of the body and due to shadowing effects from booms etc.  This effect usually proves to be a hindrance when using time intervals much longer than the spacecraft spin, and can manifest themselves in Fourier power spectra as strong peaks at the spin frequency and the associated harmonics. However for shorter time intervals it can affect the mean value, therefore it is prudent to remove it.

\begin{figure}
    \centering
    \includegraphics[angle=-90,width=0.8\textwidth]{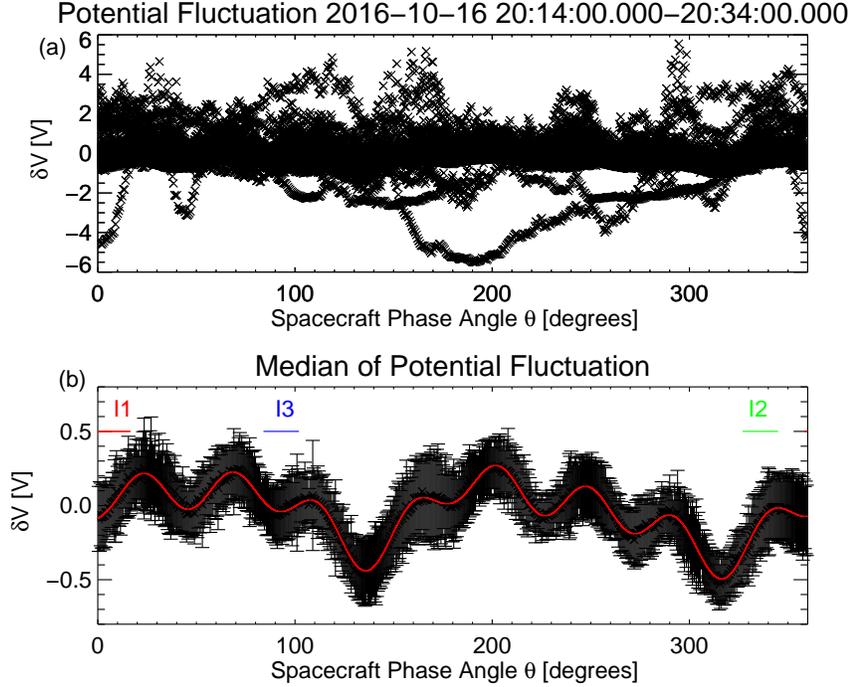}
    \caption{(a) The potential fluctuation plotted as a function of spacecraft phase angle. (b) The median value and error for each 0.5 degree bin. The red curve denotes the Model fitted to the binned data for interval 1. For reference the three coloured horizontal lines denote where in phase angle the different intervals occur.}
    \label{fig:spineffect}
\end{figure}


Spin effect removal is done by developing an empirical model of the spacecraft charging over many spins, here 20 minutes of data (60 spins) are used. The potential fluctuation is found by subtracting a local average based on a 20s spin period. The potential fluctuation data as a function of spacecraft phase angle $\theta$ is shown in Fig \ref{fig:spineffect}a. Theta is defined as $0^{\circ}$ when the Sun is in +X of the spacecraft body coordinate system. The data are then binned into spacecraft phase angle bins of half a degree and we use the median and error Eqs. \ref{median},\ref{error}. The binned data are shown in Fig \ref{fig:spineffect}b and a model $M(\theta)$ is fitted to the binned data \citep{Roberts2017} and denoted by the red curve. The model used here is a superposition of 20 sine waves and the spin effect is then subtracted from the data $V_{\text{sc sr}}=V_{\text{sc}}-M(\theta)$. Where the subscript sr denote spin removed.

It is important to note that while the spin effect is less important when considering shorter intervals it can still affect the density estimation by giving an offset value. For example if a one second interval is near zero phase angle there may not be much effect however should the phase angle be near 140$^{\circ}$ an offset of -0.5V would be expected for an interval with similar plasma parameters to interval 1. For reference the range of spacecraft phase angles for each interval are shown on Fig \ref{fig:spineffect}b as horizontal lines. 

\begin{figure}
    \centering
    \includegraphics[angle=-90,width=0.5\textwidth]{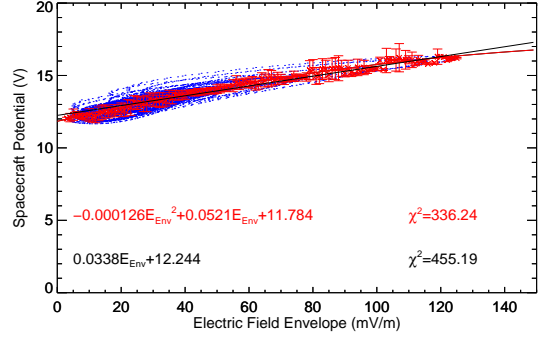}
    \caption{The relationship between the electric field envelope and the spacecraft potential. Blue points denote the measured data and red points denote the binned data. A linear fit in black and a second order polynomial (in red) are fitted to the binned data.}
    \label{fig:fig6}
\end{figure}

After the spin correction it is time to address another issue which is illustrated by the cross correlation analysis shown in Figs \ref{Lab:Fig1},\ref{Over2},\ref{Over3}. The spacecraft potential lags behind the electric field envelope. This is related to the charging properties of the spacecraft in time which can be expressed as a time constant. This is defined in Equation \ref{TRC} where $C_{sc}$ is the estimated total capacitance of the spacecraft estimated to be 2nF \citep{Graham2018}.

\begin{equation}
    \delta t_{RC}=C_{sc} \left (\frac{\partial V_{sc}}{\partial I_{\text{phot}}}\right)^{-1}
    \label{TRC}
\end{equation}

This finite charging time means that we cannot make a simple subtraction of the $E_{\text{cal}}$ from the spin corrected spacecraft potential as they would not be mapped to the correct data points due to the time lag. Additionally the spacecraft charges/discharges according to $V=V_{0}(1-e^{-t/\delta t_{RC}})$ thus to charge to $>99\%$ of the final value a time of 5 time constants need to have passed, and any fluctuations in the electric field envelope which are faster than this timescale will result in the spacecraft not reaching it's full charge. Thus at a certain frequency the spacecraft potential will no longer follow the electric field. Furthermore, as spacecraft charging occurs due to photoelectron emission, and the discharging is a result of the collection of thermal electrons. Thus the spacecraft's charging and discharging timescales can be different \citep{Wang2014}.

To correct for the time lag one approach would be to use the time lag found from the cross correlation to shift the calibrated electric field fluctuations. However, there are different charging and discharging times and the phase lag between the electric field envelope and the spacecraft potential is dependent on the timescale investigated. At low frequencies (large time scales) the Fourier coefficients are largely unaffected and higher frequencies become increasingly affected. Thus a more appropriate method is to use a spectral approach. To investigate the relationship between the electric field envelope and the spacecraft potential we calculate the wavelet coherence and phase difference \citep{Torrence1999} in Fig \ref{fig:Fig7}.

\begin{figure}
    \centering
    \includegraphics[width=0.6\textwidth]{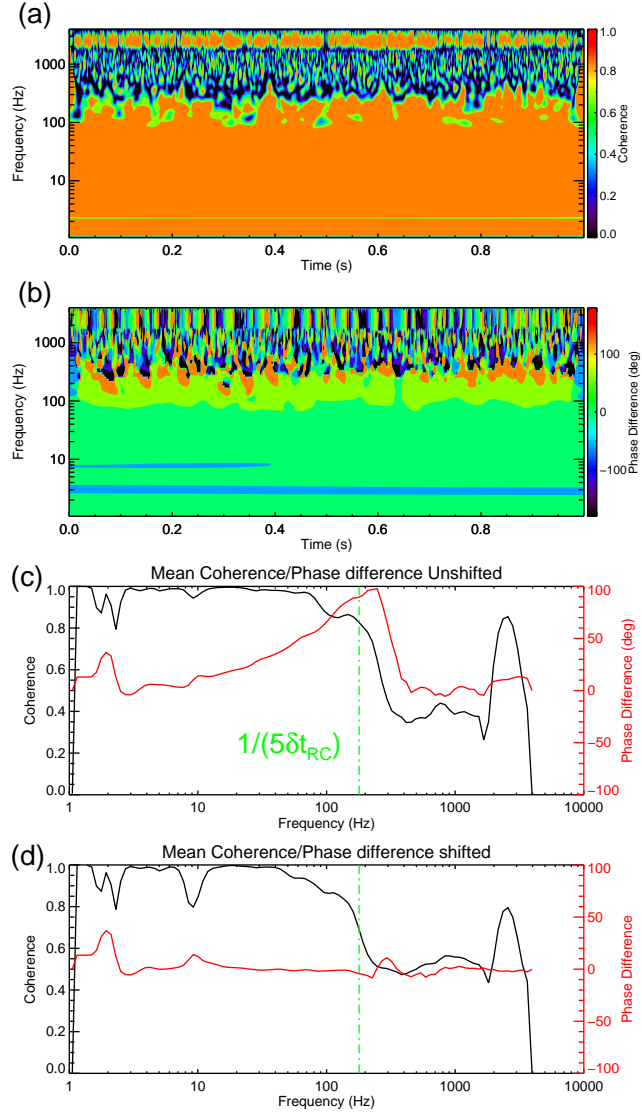}
    \caption{(a) Wavelet coherence analysis of the calibrated unshifted electric field envelope and the spacecraft potential (b) the corresponding wavelet phase difference (c) The mean coherence and phase difference (d) the mean coherence and phase difference for the shifted calibrated electric field.}
    \label{fig:Fig7}
\end{figure}

Figure \ref{fig:Fig7}a features the coherence and Fig. \ref{fig:Fig7}b shows the phase difference which are fairly constant in time. In frequency they are also constant but begin to vary near 50Hz where the phase difference begins to increase in the mean coherence and phase shown in Figure \ref{fig:Fig7}c. The rapid increase of the phase difference from 50-300Hz and decrease of the coherence here is due to the finite charging time of the spacecraft which occurs approximately near a timescale of five times the mean time constant. There is moderate decrease at around 80Hz followed by sharp decrease in coherency (Figure 9c, black curve) at 200Hz near the timescale $1/5\delta t_{RC}$, and an increase in coherency at higher frequency $\sim2000$Hz. This increase in the coherency at high frequency is a result of frequency aliasing of the Langmuir wave present (peak frequency $f=5728$Hz in the HMFE data) in the lower resolution spacecraft potential and the resampled electric field envelope. 

One interesting point to consider is why the fluctuations in the potential follow the envelope of the electric field rather than the instantaneous wave amplitude. This is likely due to the delay in the spacecraft potential's response to the electric field due to a finite charging time and a large capacitance. For solar wind the study of \cite{Chen2012b} calculates an upper estimate of 6kHz for using the spacecraft potential in the solar wind with the THEMIS spacecraft. Practically, this is limited by the sampling rate of 128Hz or instrumental noise at 100Hz when the instruments are in 'wave-burst' mode with a sampling rate of 8192Hz.  Due to the large capacitance of the spacecraft body the potential cannot respond quickly enough to the instantaneous electric field \citep[e.g.][]{Malaspina2014}. This is consistent with lower frequency waves having a larger effect on the enhanced escape i.e. that the electrons have more time to respond to the electric field and are more likely to escape. This hypothesis is supported by test particle simulations in \cite{Graham2018}, and additionally the changes in the DC electric fields cause a larger change in the potential when compared to the AC electric fields of the same magnitude \citep{Torkar2017}.

Another point to consider is that the time constant of a spacecraft may not be uniform across the spacecraft due to different surface to plasma capacitances. For example, the spacecraft body and the probes will have different time constants. Should these two capacitances be similar the response times of both components may need to be considered. The surface area of $34$m$^{2}$ for MMS corresponds to an equivalent sphere of radius 1.6m, whereas the SDP probes have a radius of 0.04m. Therefore, on the MMS spacecraft, the two capacitances are well separated. Furthermore, the bias current supplied to the probe will cause the difference in charging time-scale to be even larger. Practically this means, that only the longer timescale needs to be considered. However, this may not be the case for other spacecraft.

The coherence and phase difference properties for this event are fairly constant in time and it suggests that we can use a Fourier approach to correct for the phase lag. The phase difference between components is calculated from Eq. \ref{phi} where the tilde's denote the Fourier transform, and the Fraktur fonts denote the real and imaginary parts.

\begin{equation}
    \phi=\tan^{-1}{\Im(\tilde{E_{\text{cal}}} \tilde{V_{\text{sc sr}}}*)/\Re(\tilde{E_{\text{cal}}} \tilde{V_{\text{sc sr}}}*)}
    \label{phi}
\end{equation}

Using the obtained phase difference the Fourier coefficients of the $\tilde{E_{\text{cal}}}$ can be shifted following $\tilde{E_{\text{cal shift}}}=\tilde{E_{\text{cal}}}\exp^{-i\phi}$ which shifts the lags to zero provided they are small enough. Fourier coefficients can then be inverse transformed back to the time domain $E_{\text{cal shift}}$. This results in a relationship where the mean phase difference at all scales is low. The same coherence and phase difference analysis is performed on the shifted data and the results of the mean values are seen in \ref{fig:Fig7}d where the increasing phase difference seen in Fig \ref{fig:Fig7}c is now close to zero. Using the calibrated shifted electric field we can now subtract from the spacecraft potential to give a value of the potential where the effect of the electric field has been removed $V_{\text{sc corr}}=V_{\text{sc sr}}-E_{\text{cal shift}}$

Finally, high frequency fluctuations need to be removed where the electric field and the spacecraft potential do not follow one another and where the frequency aliasing effects can be present.  Therefore, as a final step the $V_{\text{sc corr}}$ is low pass filtered using boxcar smoothing with a window size of 10 time constants to remove these high frequency effects which occur when the coherency begins to drop near 80Hz. This value can finally be used to estimate the electron density. However, it is an important caveat to the method as well as when using the spacecraft potential more generally that there is a maximum effective time cadence due to the spacecraft charging/discharging times.

To summarize; the method proposed in this section determines the empirical relationship between the electric field envelope and the spacecraft potential, which are important in deducing the correct density values, but so far ignored in particular during the periods with high-frequency waves. This relationship is used to remove the fluctuations seen in the spacecraft potential which are due to the electric field. There is a phase difference between the electric field and the potential which needs to be corrected, and then high frequency fluctuations need to be removed. This allows the density to be estimated from the potential also when there are large electric fields, which is essential for applications where high time resolution density information are required when there are large electric fields. Some examples of potential applications would be; evaluating the Pondermotive force \citep{Henri2011} or studying electron phase space holes \cite{Tong2018,Holmes2018,Mozer2018}.

\section{Results}

\begin{figure}
    \centering
    \includegraphics[width=0.7\textwidth]{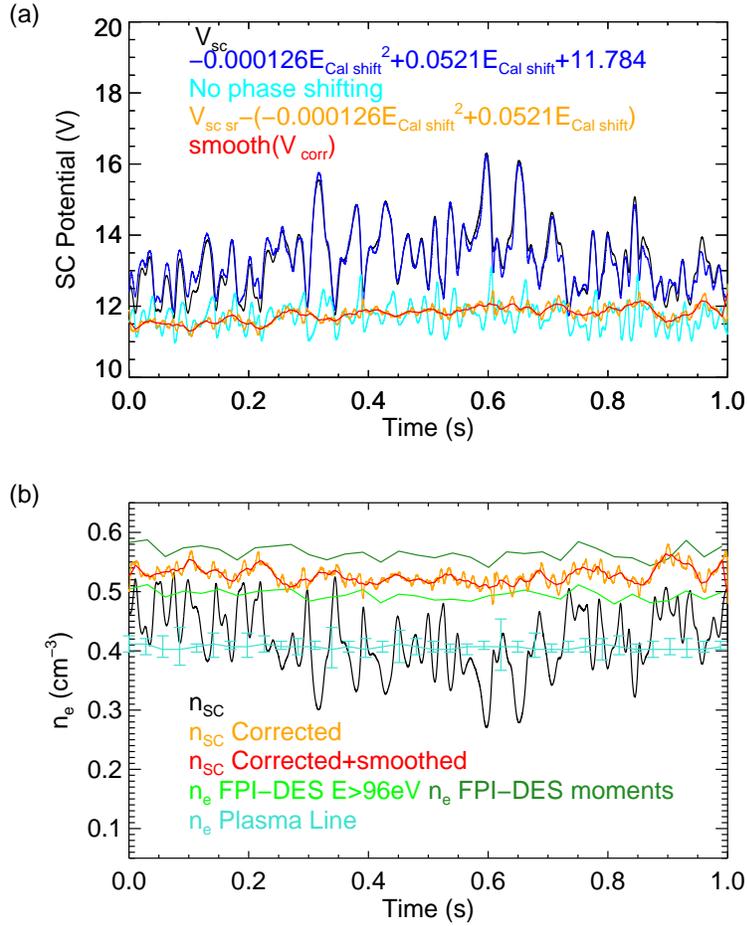}
    \caption{Results for Interval 1 (a) The original spacecraft potential is shown in black, the calibrated and phase shifted electric field in blue, the potential where fluctuations in the electric field envelope have been removed in orange, and the potential without electric field contributions which has been smoothed in red, the result of the subtraction when the phase shift is not performed is shown in cyan. (b) shows the direct measurement from FPI in green, the original estimation of the density from spacecraft potential in black, the estimations from the corresponding orange and red curves in (a). }
    \label{fig:Fig8}
\end{figure}

\begin{figure}
    \centering
    \includegraphics[width=0.7\textwidth]{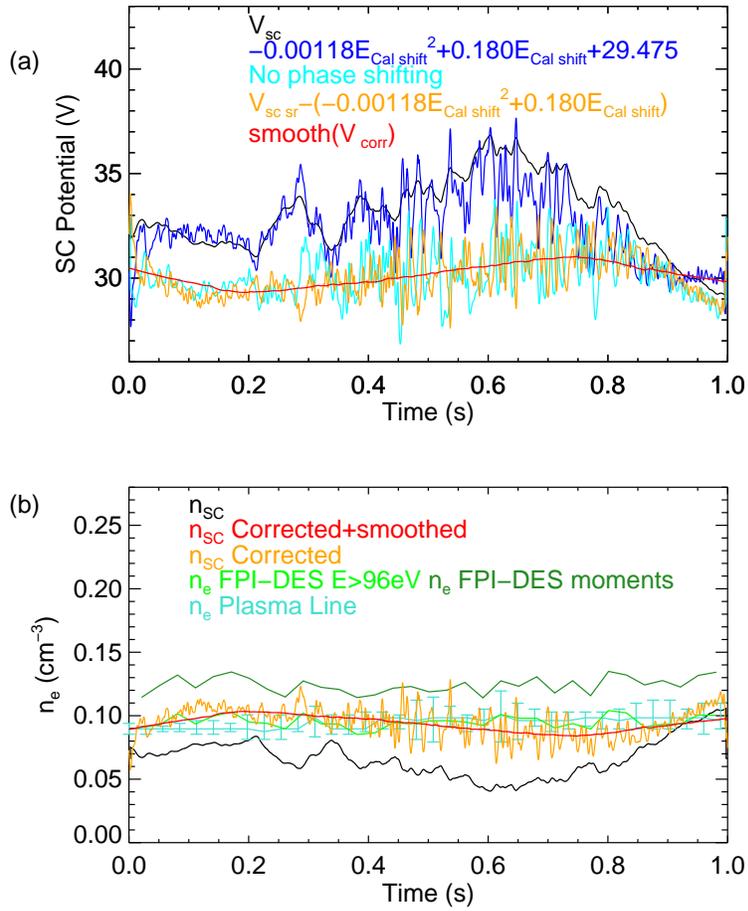}
    \caption{The same as figure \ref{fig:Fig8} but for the second interval.}
    \label{fig:Fig9}
\end{figure}

\begin{figure}
    \centering
    \includegraphics[width=0.7\textwidth]{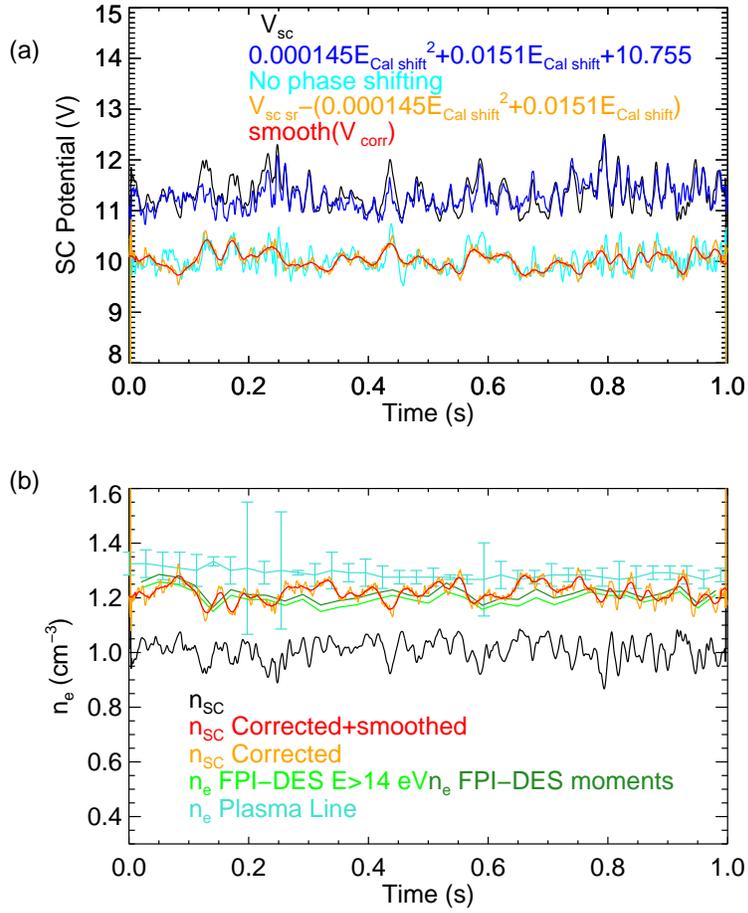}
    \caption{The same as figure \ref{fig:Fig8} but for the third interval.}
    \label{fig:Fig10}
\end{figure}

The final results for the removal of the fluctuations related to the Electric field are seen in Figs \ref{fig:Fig8}-\ref{fig:Fig10}. In the (a) panels the original spacecraft potential (not corrected for spin or electric field) is shown in black, the electric field which has been phase shifted and calibrated to volts in dark blue $E_{\text{cal shift}}$, the result when we remove the electric field fluctuation from the potential in orange and the smoothed potential in red. For comparison the cyan lines show the result when the calibrated electric field is removed without the phase shifting being performed. 

In all three cases we see that large spurious fluctuations which are a direct cause of not implementing a phase shift. This is due to the electric field leading and the spacecraft potential then following. In other words at a given time the electric field does not correspond to the potential at the same time. Thus a subtraction of these two without phase correction will give a spurious fluctuation.

Finally in Figs \ref{fig:Fig8}b - \ref{fig:Fig10}b we estimate the electron density where the colours correspond to the potentials in Figures \ref{fig:Fig8}-\ref{fig:Fig10}a. The processed data shows a clear improvement on the raw spacecraft potential data with the fluctuations due to the electric field being removed or mitigated. Figure \ref{fig:Fig8}b corresponding to the first interval shows the corrected and smoothed data showing good agreement with the electron density from FPI-DES and lies in between the moments value and the partial moments with a starting energy of 96eV. However, the measurement is somewhat different to the estimation from the plasma line which is approximately 0.15 cm$^{-3}$ smaller. This is likely due to the fact that to determine the photocurve the electron density from FPI is used, therefore it may not be surprising that it agrees best with FPI, despite the most accurate estimation coming from the plasma line. In principle the plasma line could be used to determine the photocurve with better accuracy. However, without active sounding this is not always feasible, and the natural plasma line is not always available.  The other possibility is that there is some offset due to the calibration of the electric field performed in section 3. Between the two models that were fitted there was a discrepancy of 0.5V which could introduce a systematic error of 0.05 cm$^{-3}$. Other sources of uncertainty could be in the determination of the photocurve as there is a large spread in the data points in Figure \ref{fig:2} or that the phase difference between the potential and the electric field may evolve in time and a Fourier treatment is not suitable. Finally there could be some small effects due to secondary electron emission. In this example the electron temperature is high (1keV) therefore some secondary electron emission is possible. In practice, we would expect this to be small as electron temperature is outside the peak of the secondary emission yield which is near 300-800eV \citep{Balcon2012}. However, in the same way that photoelectron escape is enhanced by acceleration from the electric field the secondary emission would also be enhanced. On average the energies of secondary electrons are higher than photoelectrons (of the order of 10 eV) \citep{Lai2011}, whereas the fitted photoelectron temperatures here are lower (see Figure \ref{fig:2}). The enhanced electron escape would therefore be more pronounced on the photoelectrons. This is because more secondary electrons can escape without needing the aid of the electric field when compared to the photoelectrons. Furthermore, as we use an empirical method relating the envelope of the electric field to the spacecraft potential both enhanced photoelectrons and enhanced secondary emission would be corrected for simultaneously.

Figure \ref{fig:Fig9} shows the result of the spacecraft potential correction the second interval which was chosen as it was an extreme event where the spacecraft potential is very large and consequently the time constant of the spacecraft is very long. The main effect of this increased time lag is that the smoothing window needs to be larger. In this extreme case $1/5\delta t_{RC} = 10$Hz meaning that although the potential can be corrected in this case and matches well with both the plasma line and the partial moments there is no advantage to using it over the partial moments. In this case the spacecraft potential is not capable of matching the time resolution of the direct measurement. Therefore in the case where the time lags are lengthy and the direct measurement with burst mode is available it would be more appropriate to use the direct measurement. However, for other spacecraft where there is not such a direct plasma measurement the density estimated from the spacecraft potential may be useful. It should also be noted that should ASPOC have been operating during this interval the smaller potential would have resulted in a smaller time delay, thus increasing the effective time resolution of the spacecraft potential density estimation. This would come at a cost of reducing the accuracy of the value of the density, due to the potential differences resulting from density fluctuations being smaller. While the focus of this work is on MMS, it is also interesting to note a different capacitance (i.e. the effect on other spacecraft) would also have an effect. A smaller capacitance would enable the potential to follow the electric field fluctuations more closely due to a reduced time constant and might even mean that the potential follows individual wave cycles rather than the envelope. In principle it might be possible to use a smaller spacecraft or even individual probe data due to their lower capacitance to test this provided that the time resolution is sufficient \citep{Malaspina2014}. Practically, the bias current to the probes ensures that the probe potential is well regulated, which would make such a measurement with a single probe difficult.

Figure \ref{fig:Fig10} shows the results from the third interval which has a lower potential and a higher density. We see that the potential is less markedly affected by the electric field i.e. only 1V contrasting with the other two cases which are affected by around 2V. This is despite the electric field having similar amplitudes in each event. This suggests that at smaller potentials the affect of the electric field is less important. In the cases where ASPOC and EDI are not operating such as those studied here smaller potentials correspond to higher densities. In Figure 4a in \cite{Graham2018} the relationship between $\frac{\delta V_{max}}{E_{max}}$ is shown to decrease with increasing density. So values of $E_{max}$ will give a smaller $\delta V_{max}$ for a larger density, suggesting this correction is more useful in low density environments. This is consistent with enhanced photoelectron escape due to the electric fields. For higher densities the electron thermal current will be much larger and changes in the photoelectron current would have a smaller effect on the potential. However, there are several instances where even small changes in the potential are important for interpretation such as inferring density fluctuations due to the pondermotive force \citep{Henri2011} or the effects of density perturbations on wave propagation  \citep[e.g.][]{Moullard2002,Yearby2011}. Therefore, the relevance of the correction presented here is also dependant on the exact use and accuracy required on $V_{sc}$ for the density estimation. Finally, for the third interval there is a large effect in the mean value due to the portion of the spacecraft which is illuminated at the given phase angle. This demonstrates the need to correct for the spin effect even when using intervals that are shorter than the spin period. In this case it is important to correct for both effects and excellent agreement is seen with the FPI-DES measurements.

\section{Summary}
In the work of \cite{Torkar2017} and \cite{Graham2018} a clear effect of the electric field on the spacecraft potential was demonstrated, causing enhanced photoelectron escape from the spacecraft. This can lead to inaccuracies in determining the electron number density from the spacecraft potential. Physically this can be interpreted as the electric field causing photoelectrons with lower energies to escape the potential well of the spacecraft. In the absence of an electric field these photoelectrons would have returned to the spacecraft. The relations that are obtained between the electric field envelope and the spacecraft potential are similar to those found in numerical simulations and in MMS data \citep{Graham2018}. There is also no difference in the electron temperature before during and after the waves ruling out trapped energetic particles causing more secondary emission. The secondary electron emission may also be enhanced, but this is small when compared to the photoemission, and the correction presented here would correct for both effects naturally. Here we have only considered the effect of the envelope of the electric field for highly electrostatic fluctuations, however it is possible that the presence of a magnetic field could also affect the photoelectron escape. A strong magnetic field or perhaps a magnetic field with a particular orientation could cause the trajectories of the photoelectrons to change due to the Lorentz force. This could cause more or fewer electrons to escape the potential well of the spacecraft \cite[e.g.][]{Malaspina2014,Wang2014a}. Numerical simulations by \cite{Graham2018} and a combination of numerical simulation and laboratory experiments \cite{Wang2014a} found that this effect becomes important for very large magnetic fields $B\sim10^3$ nT which are much larger than the typical magnetic fields at the magnetopause. Effectively the larger magnetic field causes the electron gyroradius to be smaller. When the gyroradius becomes comparable to the spacecraft size then the electrons are more likely to return to the spacecraft. These effects are likely to be small in the intervals studied here as the electric field amplitude is extreme. We also remark that the type of wave may have an effect, for example left hand circularly polarized waves may also scatter electrons leading to more electrons escaping the potential well.

To correct for the electric field effect so that the density estimation can be used we have developed a method which allows the density to be derived from the spacecraft potential even when large amplitude AC fields are present. However, there is a caveat that the correction of the phases relies on a Fourier approach, so this method may not be useful when the coherence/phase difference of electric field and spacecraft potential are not fairly constant in time, which might occur when the charging properties of the spacecraft or the plasma environment change abruptly.Furthermore the determination of the density is complicated by the charging and discharging timescales of the spacecraft, which is dependent on the potential itself. Thus in extremely sparse plasmas such as the PSBL the charging timescales can be so long that the highest frequency that can be obtained might be lower than that from FPI in burst mode data. In more dense plasmas the potential is perturbed less by the electric field than in the low density cases. Finally we have demonstrated that in three different plasma environments the electric field effect can be effectively removed and a good estimate of the mean value and the density fluctuations can still be obtained through the spacecraft potential. This is most useful when the time lags are low and the relationship between the electric field envelope and the spacecraft potential is well defined. Further work will develop this method for events where there is a time variation in the coherence properties. Additionally it is planned to investigate events when ASPOC is operating, which may make it more difficult in determining the value of density but has the advantage that the charging time is smaller.

\acknowledgments
The datasets analyzed for this study can be found in the MMS science data archive \url{https://lasp.colorado.edu/mms/sdc/public/}. Analysis of the spacecraft potential data at IWF is supported by Austrian FFG projects ASAP15/873685. Several codes for the analysis of spacecraft potential data are provided at \url{https://www.iwf.oeaw.ac.at/en/research/researchnbspgroups/space-plasma-physics/sc-plasma-interaction/mmsaspoc-data-analysis/}. O.W.R. acknowledges several helpful discussions with the FPI and ASPOC teams and Christoph Lhotka. R.N. was supported by Austrian FWF projects I2016-N20. D.B.G was supported by the Swedish National Space Board, grant 128/17. Z. V. was supported by the Austrian FWF projects P28764-N27. C.P.E. acknowledges support from the ESTEC Faculty Research Project Programme.


%
%


%
%
%
%
%

\end{document}